\documentclass{aa}
\def\Lsim{ \buildrel   <\over\sim} 
\def\lsim{\lower.75ex\hbox{$\Lsim$}}

\input psfig
\usepackage{graphics}
\usepackage{times}
\begin{document}
 
   
                                   


           \thesaurus{11     
              (11.01.2;  
               11.09.1 NGC5985;  
               11.17.3;  
               11.19.1)} 

\titlerunning{A QSO 2.4 arcsec}
\authorrunning{Halton Arp}

\title{A QSO 2.4 arcsec from a dwarf galaxy - the rest of the story}

\author{Halton Arp}
\offprints{H. Arp}
\institute{Max-Planck-Institut fuer Astrophysik\\ 85740 Garching, Germany}

\date {Received / Accepted}                

\maketitle

\begin{abstract}
D. Reimers and H.-J. Hagen report in A\&A 329, L25 (1998) 
that a QSO of $z = .807$ is a "by chance projection" on the center 
of a galaxy of $z = .009$. A considerable amount of previously 
published evidence would lead us to expect an active, low
redshift galaxy within about one degree of such a quasar. It is
shown here that, in fact, there is a bright Seyfert galaxy of 
$z = .008$ only 36.9 arcmin distance. Moreover, it turns out that a
total of five quasars, plus the dwarf galaxy, are accurately
aligned along the minor axis of this Seyfert with the quasars in 
descending order of redshift, i.e., $z = $2.13, 1.97, .81, .69 and .35.

\end{abstract}

\section{Introduction}
A very blue galaxy, SBS 1543+593, was discovered in the second 
Byurakan survey (Markarian et al. 1986). The Hamburg objective 
prism survey (Hagen et al. 1995) rediscovered it as a quasar and 
subsequent spectra and imaging revealed it to be a quasar of $z =
.807$ only 2.4 arcsec from the center of a galaxy of dwarf
characteristics (Reimers and Hagen 1998).

     Despite computing the probability of chance proximity as 
 $1.5\times10^{-3}$, Reimers and Hagen stated that the center of 
the galaxy was a chance projection on a background quasar. It was 
not taken into account, however, that large, and 
statistically significant, numbers of quasars fell very close to 
many other low redshift galaxies (for lists of cases see 
G.R.Burbidge 1996). Moreover, Reimers and Hagen themselves 
estimated that the QSO was very little reddened. Of course if 
it was behind the inner part of the dwarf spiral it should have been noticeably reddened. 
\footnote{For $S_d$ spirals Sandage and Tamman (1981) use mean absorptions of 
  .28 to .87 mag. ranging from face-on to edge-on systems.  For a quasar
  entirely behind the galaxy this would give $.6 \lsim A_B \lsim 1.7$ mag. and 
  $.15 \lsim E_{B-V} \lsim .44$mag.  However the quasar here would have to 
shine
  through a region very close to the center where the absorption is much higher
  than average.  For example, Hummel, van der Hulst and Keel (1987) show
  absorptions to the center of the late type spiral NGC 1097 to be $.3 \lsim
  A_v \lsim 3.0$mag.  That would translate to a full reddening of $.2 \lsim
  E_{B-V} \lsim 2.0$mag.  which again would be conspicuous.}

     The dwarf spiral, 
however, did not seem like a galaxy with an active nucleus  which 
would eject a quasar. {\it Therefore I asked the question: "Is there
a nearby galaxy which could be the origin of the quasar?"} It
quickly turned out that only 36.9 arcmin distant was the very
bright (V = 11.0 mag.) Seyfert galaxy, NGC5985. This  was just
the distance with\-in which quasars were found associated at a 7.4 sigma
 significance  with a 
large sample of Seyferts of similar apparent magnitude (Radecke 
1997; Arp 1997a). The following sections present the full census 
of objects around the Seyfert NGC5985. 

\section{The quasars around the Seyfert NGC5985}   

\begin{figure*}
\mbox{\psfig{figure=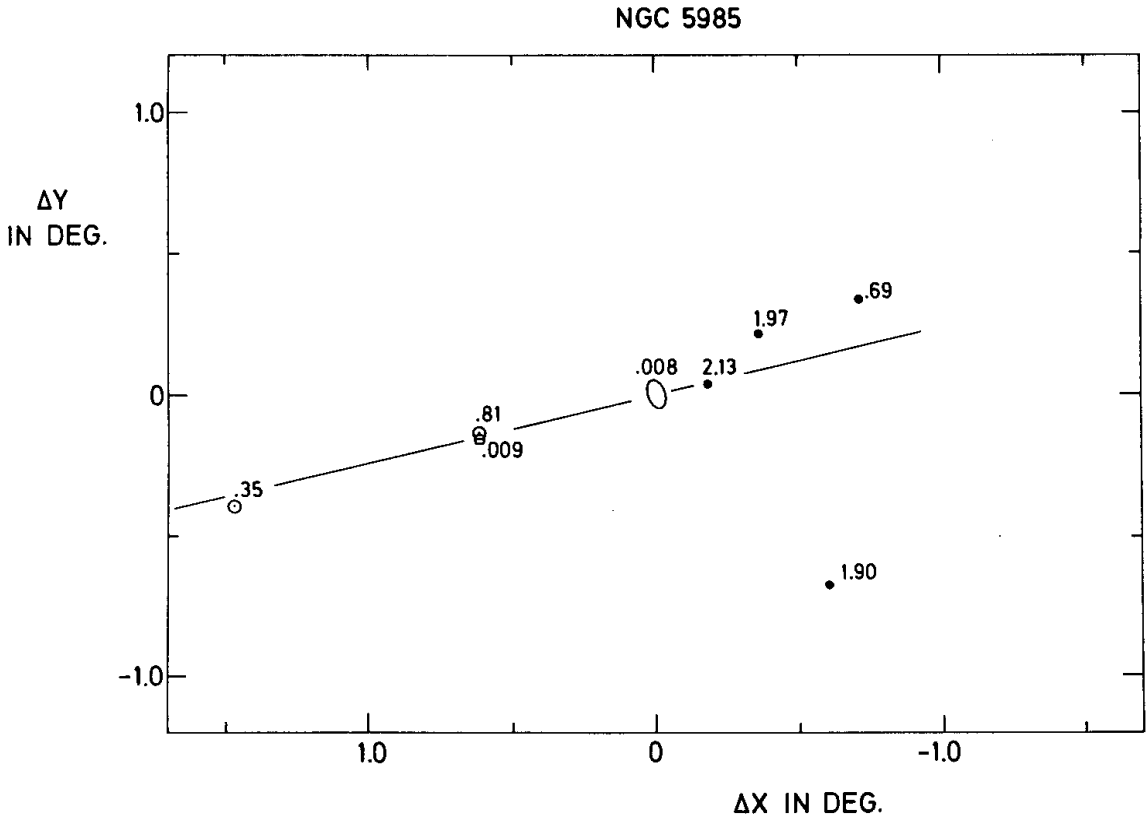,height=11.6cm}}
\caption[]{All catalogued active galaxies and quasars within the
pictured area are plotted. Redshifts are labeled. The dwarf
spiral 2.4 arc sec from the z = .81 quasar has a z = .009 which
marks it as a companion of the Seyfert NGC5985 at z = .008. The 
line represents the position of the minor axis of NGC5985.}
\label{fig1}
\end{figure*}

Fig. 1 shows a plot of all the catalogued active galaxies
(V\'eron and V\'eron 1996) in an area 3.4 x 2.4 degrees centered on
NGC\-5985. It is apparent that there is only one active galaxy in 
the area and it is the bright, Shapley-Ames Seyfert, NGC5985.
The chance of finding a Seyfert galaxy this bright, this close to the $z =
.81$ quasar is $< 10^{-3}$.

One caveat is that if the $z = .009$ galaxy is lensing the $z  = .81$ QSO then this 
probability will increase because of the tendency of galaxies to cluster. We have, 
however,  argued against the lensing hypothesis on the basis that the QSO appears 
unreddened. Reimers and Hagen also have suggested that the central mass of the galaxy 
may be too small for lensing to be important.

All catalogued quasars, plus the new QSO of $z = .807$ and its companion are also 
plotted. It turned out all quasars are from a uniformly searched region in the second Byurakan survey. 
Since five of the six quasars are aligned within about $\pm 15^\circ$ accuracy one can compute the 
probability of their accidental alignment as $P^5_6 (\leq 15^\circ) = 6 \times 10^{-4}$.
 But then one must factor in
the probability that the line would agree with an {\it a priori} minor axis
direction.  This would reduce the probability by another factor of $10^{-2}$.
The total probability of accidentally finding this prototypical configuration
is then of the order of $10^{-8}$ to $10^{-9}$.



\begin{table*}
\begin{center}
\caption{Objects in the field around NGC5985}
\begin{tabular}{|l|c|l|r|r|r|r|r|r|}
\hline
Object&Type&Vmag.&$z$&R.A.(2000)&\llap{D}ec.\hfill&$\Delta x^\circ$&
$\Delta y^\circ $&$r^\prime$\\\hline
NGC5985&Sey1&10.98&.008&15/39/37.5&{+}59/19/58
&---&---&---\\ 
HS 1543+5921&QSO&16.4&.807&15/44/20.1&59/12/26&.6019&-.1256&36.9\\
SBS 1543+593&dSp&16.9&.009&$\prime\prime$\quad \ &24&$\prime\prime$\ \ \ 
 & -.1262&$\prime\prime$ \ \ \\
RXJ 15509+5856&S1&16.4&.348&15/50/56.8&58/56/04&1.4535&-.3984&90.4\\
SBS1537+595&QSO&19.0&2.125&15/38/06.0&59/22/36&-.1944&+.0439&12.0\\SBS1535+596&QSO&19.0&1.968&15/36/45.7&59/32/33&-.3644&+.2097&25.2\\
SBS1533+588&QSO&19~~&1.895&15/34/57.2&58/39/24&-.6016&-.6761&54.3\\
SBS1532+598&QSO&18.5&.690&15/33/52.8&59/40/19&-.729&+.339&48.2\\
\hline
\end{tabular}
\end{center}
\end{table*}

Whether one considers the $z = 1.90$ quasar an unrelated interloper or an
ejection in a different direction does not change the calculation appreciably.

     The parameters of the objects in this field are listed in
Table 1. Their distances from NGC5985, at the center of the
field, are calculated in degrees.

\section{Similar associations which have been previously 
published}
From 1966 onward, evidence that quasars were associated with 
low redshift galaxies has been presented (for review see Arp 
1987). Since the discovery of a pair of quasars across the 
Seyfert NGC4258 (Pietsch et al. 1994; E.M.Burbidge 1995), 
however, bright Seyfert galaxies have been systematically 
analyzed\break (Radecke 1997; Arp 1997a). The latter investigations 
demonstrated physical associations at greater than 7.5 sigma for 
qua\-sars within about $10' < r < 40'$ of these active galaxies. 

     The alignment of quasars shown in Fig. 1 exhibits the same
properties found in the previously published results except that 
the scale of the separations and apparent magnitudes of the
qua\-sars in the NGC5985 system suggest it may be somewhat closer 
than the average Seyfert previously investigated. In the NGC\-5985
association, however, there are more than the usual pair of 
quasars so that the line is very well defined. In this respect 
it is another crucial case like the 6 quasars associated with 
NGC3516 (Chu et al. 1998). The NGC3516 case is shown here in 
Fig. 2 for comparison.

\begin{figure}
\mbox{\psfig{figure=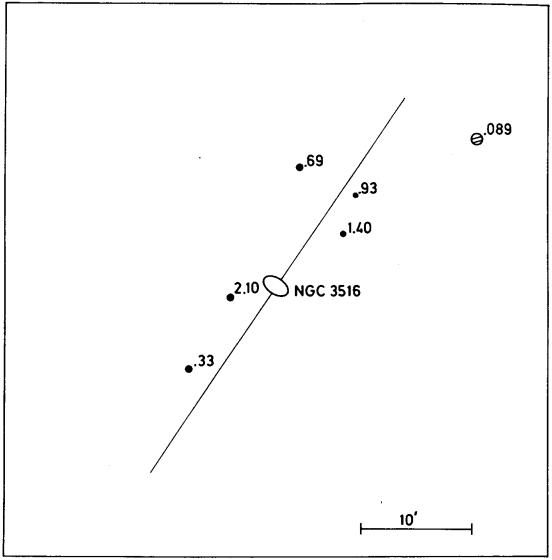,height=8.1cm}}
\caption[]{ All the X-ray bright quasars around the bright Seyfert
NGC3516 are plotted with their redshifts labeled. (From Chu et
al. 1998). The line represents the direction of the galaxy minor 
axis.}
\label{fig2}
\end{figure}

\section{Alignment along the minor axis}
Early in the association of quasars with active galaxies it was
noticed that there was a tendency for them to lie in the
direction of the minor axis of the galaxy. When the X-ray pairs 
began to be identified this correlation strengthened (Arp 1997c; 1998a,b). 
Particularly in Arp (1998a,b) it was shown that
the quasars associated with Seyferts with well defined minor axes fell along
this minor axis within a cone of about $\pm 20^\circ$ opening angle.  These
same references showed companion galaxies, as in the present case of NGC 5985,
also preferentially falling along this same axis.
 With the observations of NGC3516 shown in 
Fig. 2, however, there appeared a continuous definition of the 
minor axis by the quasars. 

     The most striking aspect of the NGC5895 case now becomes
the fact that the line in Fig. 1 was not drawn through the
quasars as one might assume! The line in Fig. 1 actually plots 
the direction of the minor axis of the Seyfert as recorded in 
the Uppsala General Catalogue of Galaxies (Nilson 1973). 

     There are two important conclusions to be drawn: The first 
is that the quasars must originate in an ejection process. The 
minor axis is the expected ejection direction from an active 
galaxy nucleus (see also the discussion of ejection along the 
minor axis in NGC4258 in Burbidge and Burbidge 1997). The second 
is that the chance of the observed quasars falling so closely 
along a {\it predicted} line by accident is negligible, thus 
confirming at an extraordinarily high level of significance the 
physical association of the quasars with the low redshift 
galaxy. 

\section{The role of companion galaxies}
What is the origin of the dwarfish spiral only 2.4 arcsec from
the $z = .81$ quasar? The simplest answer is that it represents
entrained material from NGC5985 which accompanied the ejection of
the $z = .81$ quasar. The redshift of NGC5985 is $z = .008$ and the
redshift of the dwarf spiral is $z = .009$. The latter redshift
and its distance from NGC5985 would pretty clearly identify it
as a physical companion of the Seyfert.
 
     It has been suggested that dwarfs are associated with
ejection from a central galaxy and an example is the string of
dwarfs coming from the X-ray ejecting NGC4651 (Arp 1996, Fig.6).
But, in general, companion galaxies have been identified since
1969 as falling preferentially along the minor axes of their
central galaxy. This has been interpreted as high redshift
quasars evolving to lower redshifts and finally into companion
galaxies (Arp 1997b; 1998a,b). The line of ejection seems 
to be remarkably stable from NGC5985 so objects of older 
evolutionary age could be going out or falling in along the same 
track. This would give a much higher chance of the quasar and 
the dwarf being accidently nearby at any time. In fact, observing 
quasars and galaxies lying along the same ejection lines from active 
galaxies gives for the first time an explanation for the many cases 
of close associations of higher redshift quasars with low redshift galaxies
 (G. Burbidge 1996) and higher redshift quasars with lower redshift quasars
(Burbidge, Hoyle and Schneider 1997). {\it I also note the presence of three
NGC 
companion galaxies closely along NGC5985's minor axis to the NW. These 
are of fainter apparent magnitude and one of them has a measured redshift 
(z = .0095), similar to the dwarf SE in that it has a few hundreds of km/sec 
higher redshift than the parent Seyfert. This is just as had been found for
 the redshift behavior of companion galaxies (see Arp 1998a,b).}

\section{Intrinsic redshift vs angular separation from the galaxy}
Since the quasars are strictly ordered in decreasing redshift
with distance from the central galaxy in NGC5985, it is 
interesting to compare the relation with the one in NGC3516 
where they are also so ordered. Fig. 3 show several interesting 
results:

\begin{figure}
\mbox{\psfig{figure=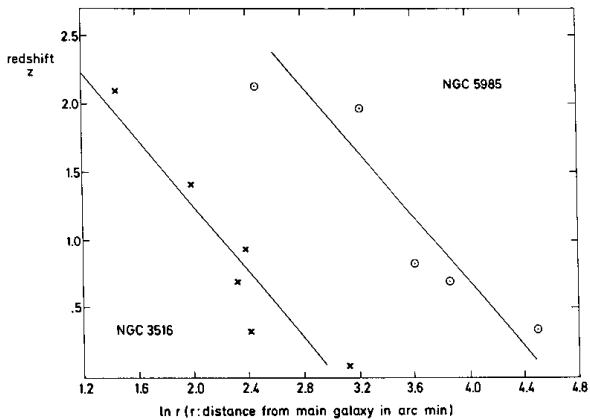,height=5.8cm}}
\caption[]{The relation between the redshift of the quasars and their 
distance from the active galaxy for the two best cases of 
multiple, aligned quasars. The exponential law is indicated to
be similar but the scale of the NGC5985 relation is larger by a 
factor of about 4.5, indicating NGC5985 is nearer the observer 
and/or oriented more  across the line of sight.}
\label{fig3}
\end{figure}

\begin{itemize}
\item  The slopes of the relation in both systems are closely
the same. Since the plot is in ln r, this means the relationship
is exponential with the same exponent in both cases. 
\item The constant separation between the two slopes translates
into a difference in scale between the two systems of a factor
of about 4.5. 
\end{itemize}

     The latter scale factor could be accounted for if the
NGC5985 system were closer to the observer than the NGC3516
system. What do the apparent magnitudes suggest? The central
Seyfert NGC5985, in dereddened magnitudes, is about 1.33 mag.
bright\-er than NGC3516. The apparent magnitudes of the quasars in
the NGC3516 association have not been accurately measured but
the quasars in the NGC5985 system appear to be 2 to 2.5 mags.
brighter than quasars measured around Seyferts generally at the 
distance of the Local Supercluster. If we therefore estimate
that NGC5985 is at a distance modulus about 2 mag. closer than
NGC3516 then the scale of the alignment should be about 2.5
times greater. Since the projected ellipticities of the two
galaxies suggests the minor axis of  NGC3516 is oriented at least 
45\% closer to the line of sight, a total scale difference of 3.6
out of 4.5 is accounted for.\footnote{The inclination of NGC 5985
 is taken from de Vaucouleurs et al. (1976) and of NGC3516 from Arribas et al. (1997).}
 It seems quite possible that the
separation of the quasars from the galaxies as a function of
intrinsic redshift is quite similar. This is an important result
to refine because it implies, for similar ejection velocities,
that the evolution rate to lower redshifts with time is a
physical constant. 

     One further comment concerns the quasars near $z = 2$. As
found in previous associations they are more than 2 magnitudes
fainter than the $z \approx 0.3$ to $1.4$ quasars. This predicts that for
the majority of Seyferts which have medium redshift quasars in
the 18 to 19th mag. range, that their $z = 2$ quasars should be
found in the 20 to 21+ mag. range and closer to their active
galaxy of origin. These quasars, generally weak in X-rays,
should be searched for with grism detectors on large aperture
telescopes such as was done for M82 (E.M. Burbidge et al. 1980). 

\section{Quantization of redshifts}
NGC3516 was unusual in that each of the six quasars fell very
close to each of the major redshift peaks observed for quasars
in general. For radio selected quasars with $S_{11} > 1$Jy it was shown (Arp et al. 1990)
 that the Karlsson formula 
$$(1 + z_2) = (1 + z_1) 1.23$$
$$z_n  =  .061, .30, .60, .96, 1.41, 1.96, 2.64 \dots$$
was fitted with a confidence level of 99 to 99.97\%. In  the present NGC5985
case, four of the five quasars fall close to the quantized
redshift values. Therefore, together with NGC 3516, in the two best cases of multiple,
aligned quasars the redshifts fall very close to the formula
peaks in 10 out of 11 cases. The z = .81 redshift then may
represent a short-lived phase in evolution from the .96 to the
.60 peak.

\section{Summary}
Examination of catalogued objects in the vicinity of a QSO
only 2.4 arcsec from a dwarf galaxy reveal that this pair of
objects, as well as four additional quasars, are associated
with a nearby, bright Seyfert galaxy.The configuration satisfies
all the criteria of previous quasar - active galaxy
associations. In particular the NGC5985 alignment of five quasars
defines well the minor axis of the galaxy as does the alignment
of six quasars through the Sey\-fert NGC3516. The similar decline
in quantized redshift values with increasing distance from the
active galaxy suggests there is one law of redshift evolution
with time.

\end{document}